# Time Warp Edit Distance

**PIERRE-FRANÇOIS MARTEAU**
pierre-francois.marteau@univ-ubs.fr

**VALORIA, UNIVERSITE EUROPEENNE DE BRETAGNE**
CAMPUS DE TOHANNIC, BAT. YVES COPPENS, BP 573, 56017 VANNES CEDEX, FRANCE
**FEBRUARY 2008**

**Abstract**: This technical report details a family of time warp distances on the set of discrete time series. This family is constructed as an editing distance whose elementary operations apply on linear segments. A specific parameter allows controlling the stiffness of the elastic matching. It is well suited for the processing of event data for which each data sample is associated with a timestamp, not necessarily obtained according to a constant sampling rate. Some properties verified by these distances are proposed and proved in this report.

# I. Introduction

At the junction of symbolic edit distances [1][3][5][7][11] and dynamic time warping measures [2][9][4][10] we propose a family of Time Warp Edit Distance (TWED) that we denote $\delta_{\lambda,\gamma}$ to refer to the two parameters that characterize the family, namely the gap penalty $\lambda$ and the stiffness parameter $\gamma$. We first define $\delta_{\lambda,\gamma}$ and then we prove successively that whenever $\lambda \geq 0, \gamma > 0$

1. <u>Proposition 1</u>: $\delta_{\lambda,\gamma}$ is a distance metric.

2. <u>Proposition 2</u>: $\delta_{\lambda,\gamma}$ is upper bounded by twice the *L1* distance.

3. <u>Proposition 3</u>: $\delta_{\lambda,\gamma}$ is an increasing function of $\lambda$ and $\gamma$.

4. <u>Proposition 4</u>: Upper-bounding the distance between a time series and its piecewise constant polygonal approximation

Further details and experiments on $\delta_{\lambda,\gamma}$ are described in [6].

## II. Definitions

Let $U$ be the set of finite discrete time series: $U = \{A_1^p / p \in N^+\} \cup \{\Omega\}$, where $\Omega$ is the empty time series (with null length). Let $A_1^p$ be a time series with discrete index varying between $1$ and $p$. Let $a'_i$ be the $i^{th}$ sample of time series $A$. We will consider that $a'_i \in S \times T$ where $S \subset R^k$ with $k \geq 1$ embeds the multidimensional space variables and $T \subset R$ embeds the time stamp variable, so that we can write $a'_i = (a_i, t_{a_i})$ where $a_i \in S$ and $t_{a_i} \in T$, with the condition that $t_{a_i} > t_{a_j}$ whenever $i > j$ (time stamp are strictly increasing in the sequence of samples).

Let us define $\delta_{\lambda,\gamma}$ as:

$$\delta_{\lambda,\gamma}(A_1^p, B_1^q) = Min \begin{cases} \delta_{\lambda,\gamma}(A_1^{p-1}, B_1^q) + \Gamma(a'_p \to \Lambda) \\ \delta_{\lambda,\gamma}(A_1^{p-1}, B_1^{q-1}) + \Gamma(a'_p \to b'_q) \\ \delta_{\lambda,\gamma}(A_1^p, B_1^{q-1}) + \Gamma(\Lambda \to b'_q) \end{cases} \quad (1)$$

$$\text{with} \quad \begin{aligned} \Gamma(a'_p \to \Lambda) &= d(a'_p, a'_{p-1}) + \lambda \\ \Gamma(a'_p \to b'_q) &= d(a'_p, b'_q) + d(a'_{p-1}, b'_{q-1}) \\ \Gamma(\Lambda \to b'_q) &= d(b'_{q-1}, b'_q) + \lambda \end{aligned}$$

where $d$ is any distance on $R^{k+1}$. In practice, we will choose $d(a', b') = d_{LP}(a, b) + \gamma . d_{Lp}(t_a, t_b)$ where $\gamma$ is a parameter which characterizes the *stiffness* of the elastic distance $\delta_{\lambda,\gamma}$, and $\lambda$ any positive constant element in $R$ that corresponds to a gap penalty.

The recursion is initialized setting:

$$\delta_{\lambda,\gamma}(A_1^0, B_1^0) = 0$$

$$\delta_{\lambda,\gamma}(A_1^0, B_1^j) = \sum_{k=1}^{j} d(b'_k, b'_{k-1}), j \in \{1,...,q\}$$

$$\delta_{\lambda,\gamma}(A_1^i, B_1^0) = \sum_{k=1}^{i} d(a'_k, a'_{k-1}), i \in \{1,...,p\}$$

with $a'_0 = b'_0 = 0$ by convention.

## III. Proposition 1: $\delta_{\lambda,\gamma}$ is a distance on the set *U* of finite discrete time series

*Proof:*

*P1: non-negativity*

For all $(A_1^p, B_1^q)$ in $U \times U$ let $m=p+q$. Non-negativity of $\delta_{\lambda,\gamma}$ is proved by induction on *m*.

*P1* is true for *m=0* by definition of $\delta_{\lambda,\gamma}$ and the induction hypothesis holds.

Suppose *P1* is true for all $m \in \{0,..n-1\}$ for some *n>0*. Then for all $(A_1^p, B_1^q)$ in $U \times U$ such that $m = n$, as $\delta_{\lambda,\gamma}(A_1^{p-1}, B_1^q), \delta_{\lambda,\gamma}(A_1^{p-1}, B_1^{q-1})$ and $\delta_{\lambda,\gamma}(A_1^p, B_1^{q-1})$ are assumed positive and as the non-negativity of distance *d* holds, $\delta_{\lambda,\gamma}(A_1^p, B_1^q)$ is necessary non-negative, showing that P1 is true for all $m \in \{0,..n\}$. By induction, *P1* holds for all $m \in N$. □

## P2: identity of indiscernibles

For all $(A_1^p, B_1^q)$ in $U \times U$, if $A_1^p = B_1^q$ then $p=q$ and $\forall i \in \{1,...,p\}, a'_i = b'_i$. It is easy to show by induction on $p$ that if $A_1^p = B_1^q$ then $0 \leq \delta_{\lambda,\gamma}(A_1^p, B_1^q) \leq \sum_{i=1}^{p} d(a'_i, b'_i) = 0$ leading to $\delta_{\lambda,\gamma}(A_1^p, B_1^q) = 0$.

Now consider the backward proposition P'2: $\delta_{\lambda,\gamma}(A_1^p, B_1^q) = 0 \Rightarrow A_1^p = B_1^q$. P'2 is proved by induction on $m=p+q$.

P'2 is true for m=0.

Suppose P'2 is true for all $m \in \{0,..n-1\}$ for some $n>0$. Then for all $(A_1^p, B_1^q)$ in $U \times U$ such that $m = n$ and $\delta_{\lambda,\gamma}(A_1^p, B_1^q) = 0$ we have necessarily:

$$\delta_{\lambda,\gamma}(A_1^p, B_1^q) = \delta_{\lambda,\gamma}(A_1^{p-1}, B_1^{q-1}) + d(a'_p, b'_q) + d(a'_{p-1}, b'_{q-1}).$$

We verify that the cases where $\delta_{\lambda,\gamma}(A_1^p, B_1^q) = \delta_{\lambda,\gamma}(A_1^p, B_1^{q-1}) + d(b'_q, b'_{q-1}) + \lambda$ or $\delta_{\lambda,\gamma}(A_1^p, B_1^q) = \delta_{\lambda,\gamma}(A_1^{p-1}, B_1^q) + d(a'_p, a'_{p-1}) + \lambda$ are impossible since $d(b'_q, b'_{q-1})$ and $d(a'_p, a'_{p-1})$ are strictly positive (the reason being that time stamps are strictly increasing). Thus, $\delta_{\lambda,\gamma}(A_1^{p-1}, B_1^{q-1}) = 0$ and $d(a'_p, b'_q) + d(a'_{p-1}, b'_{q-1}) = 0$ leading to $A_1^{p-1} = B_1^{q-1}$ and $a'_p = b'_q$. Finally $p=q$ and necessarily $A_1^p = B_1^q$ $\square$.

## P3: Symmetry

Proof: Since the distance $d$ on the sample space $S \times T$ is symmetric, it is easy to show that $\delta_{\lambda,\gamma}(A_1^p, B_1^q)$ is symmetric for all $(A_1^p, B_1^q)$ in $U \times U$ by induction on $m=p+q$. $\square$

## P4: Triangle inequality

For all $(A_1^p, B_1^q, C_1^r)$ in $U \times U \times U$, $\delta_{\lambda,\gamma}(A_1^p, C_1^r) \leq \delta_{\lambda,\gamma}(A_1^p, B_1^q) + \delta_{\lambda,\gamma}(B_1^q, C_1^r)$.

Proof: We will prove *P4* by induction on *m=p+q+r*.

*P4* is true for *m=0* since $\delta_{\lambda,\gamma}(\Omega,\Omega) = 0 \leq \delta_{\lambda,\gamma}(\Omega,\Omega) + \delta_{\lambda,\gamma}(\Omega,\Omega)$ and the induction hypothesis holds.

*(H4):* Suppose *P4* is true for all $m \in \{0,..n-1\}$ for some *n>0*. Let $\Sigma = \delta_{\lambda,\gamma}(A_1^p, B_1^q) + \delta_{\lambda,\gamma}(B_1^q, C_1^r)$. Then for all $(A_1^p, B_1^q, C_1^r)$ in $U \times U \times U$ such that *m=n*, we have basically 9 different cases to explore for the decomposition of $\delta_{\lambda,\gamma}(A_1^p, B_1^q)$ and $\delta_{\lambda,\gamma}(B_1^q, C_1^r)$:

**1st Case:** if
$$\begin{cases} \delta_{\lambda,\gamma}(A_1^p, B_1^q) = \delta_{\lambda,\gamma}(A_1^{p-1}, B_1^{q-1}) + \Gamma(a'_p \to b'_q) \\ \delta_{\lambda,\gamma}(B_1^q, C_1^r) = \delta_{\lambda,\gamma}(B_1^{q-1}, C_1^{r-1}) + \Gamma(b'_q \to c'_r) \end{cases}, \text{then}$$

$\Sigma \geq \delta_{\lambda,\gamma}(A_1^{p-1}, B_1^{q-1}) + \delta_{\lambda,\gamma}(B_1^{q-1}, C_1^{r-1}) + d(a'_p, b'_q) + d(a'_{p-1}, b'_{q-1}) + d(b'_q, c'_r) + d(b'_{q-1}, c'_{r-1})$

$\geq \delta_{\lambda,\gamma}(A_1^{p-1}, C_1^{r-1}) + d(a'_p, c'_r) + d(a'_{p-1}, c'_{r-1}) \underset{def}{\geq} \delta_{\lambda,\gamma}(A_1^p, C_1^r)$  since (*H4*) applies,

and *d* verifies the triangular inequality.

**2nd Case:** if
$$\begin{cases} \delta_{\lambda,\gamma}(A_1^p, B_1^q) = \delta_{\lambda,\gamma}(A_1^p, B_1^{q-1}) + \Gamma(\Lambda \to b'_q) \\ \delta_{\lambda,\gamma} B_1^q, C_1^r) = \delta_{\lambda,\gamma}(B_1^{q-1}, C_1^{r-1}) + \Gamma(b'_q \to c'_r) \end{cases}, \text{then}$$

$$\Sigma = \delta_{\lambda,\gamma}(A_1^p, B_1^{q-1}) + d(b'_q, b'_{q-1}) + \lambda + \delta_{\lambda,\gamma}(B_1^{q-1}, C_1^{r-1}) + d(b'_q, c'_r) + d(b'_{q-1}, c'_{r-1})$$

$$\geq \delta_{\lambda,\gamma}(A_1^p, B_1^{q-1}) + \delta_{\lambda,\gamma}(B_1^{q-1}, C_1^{r-1}) + d(c'_r, c'_{r-1}) + \lambda, \quad (d \text{ verifies the triangular inequality})$$

$$\geq \delta_{\lambda,\gamma}(A_1^p, C_1^{r-1}) + d(c'_r, c'_{r-1}) + \lambda, \quad (H4) \text{ applies}$$

$$\underset{def}{\geq} \delta_{\lambda,\gamma}(A_1^p, C_1^r)$$

**3$^{rd}$ Case:** if $\begin{cases} \delta_{\lambda,\gamma}(A_1^p, B_1^q) = \delta_{\lambda,\gamma}(A_1^{p-1}, B_1^q) + \Gamma(a'_p \to \Lambda) \\ \delta_{\lambda,\gamma}(B_1^q, C_1^r) = \delta_{\lambda,\gamma}(B_1^{q-1}, C_1^{r-1}) + \Gamma(b'_q \to c'_r) \end{cases}$, then

$$\Sigma = \delta_{\lambda,\gamma}(A_1^{p-1}, B_1^q) + d(a'_p, a'_{p-1}) + \lambda + \delta_{\lambda,\gamma}(B_1^q, C_1^r)$$

$$\Sigma \geq \delta_{\lambda,\gamma}(A_1^{p-1}, C_1^r) + d(a'_p, a'_{p-1}) + \lambda \quad (H4) \text{ applies}$$

$$\underset{def}{\geq} \delta_{\lambda,\gamma}(A_1^p, C_1^r)$$

**4$^{th}$ Case:** if $\begin{cases} \delta_{\lambda,\gamma}(A_1^p, B_1^q) = \delta_{\lambda,\gamma}(A_1^p, B_1^{q-1}) + \Gamma(\Lambda \to b'_q) \\ \delta_{\lambda,\gamma}(B_1^q, C_1^r) = \delta_{\lambda,\gamma}(B_1^q, C_1^{r-1}) + \Gamma(\Lambda \to c'_r) \end{cases}$, then

$$\Sigma = \delta_{\lambda,\gamma}(A_1^p, B_1^q) + \delta_{\lambda,\gamma}(B_1^q, C_1^{r-1}) + d(c'_r, c'_{r-1}) + \lambda$$

$$\Sigma \geq \delta_{\lambda,\gamma}(A_1^p, C_1^{r-1}) + d(c'_r, c'_{r-1}) + \lambda \quad (H4) \text{ applies}$$

$$\underset{def}{\geq} \delta_{\lambda,\gamma}(A_1^p, C_1^r)$$

**5$^{th}$ Case:** if $\begin{cases} \delta_{\lambda,\gamma}(A_1^p, B_1^q) = \delta_{\lambda,\gamma}(A_1^{p-1}, B_1^{q-1}) + \Gamma(a'_p \to b'_q) \\ \delta_{\lambda,\gamma}(B_1^q, C_1^r) = \delta_{\lambda,\gamma}(B_1^q, C_1^{r-1}) + \Gamma(\Lambda \to c'_r) \end{cases}$, then

$$\Sigma = \delta_{\lambda,\gamma}(A_1^p, B_1^q) + \delta_{\lambda,\gamma}(B_1^q, C_1^{r-1}) + d(c'_r, c'_{r-1}) + \lambda$$

$$\geq \delta_{\lambda,\gamma}(A_1^p, C_1^{r-1}) + +d(c'_r, c'_{r-1}) + \lambda \quad (H4) \text{ applies}$$

$$\underset{def}{\geq} \delta_{\lambda,\gamma}(A_1^p, C_1^r)$$

**6th Case**: if $\begin{cases} \delta_{\lambda,\gamma}(A_1^p, B_1^q) = \delta_{\lambda,\gamma}(A_1^{p-1}, B_1^q) + \Gamma(a'_p \to \Lambda) \\ \delta_{\lambda,\gamma}(B_1^q, C_1^r) = \delta_{\lambda,\gamma}(B_1^q, C_1^{r-1}) + \Gamma(\Lambda \to c'_r) \end{cases}$, then

$$\Sigma = \delta_{\lambda,\gamma}(A_1^p, B_1^q) + \delta_{\lambda,\gamma}(B_1^q, C_1^{r-1}) + d(c'_r, c'_{r-1}) + \lambda$$
$$\geq \delta_{\lambda,\gamma}(A_1^p, C_1^{r-1}) + + d(c'_r, c'_{r-1}) + \lambda \qquad (H4) \text{ applies}$$
$$\underset{def}{\geq} \delta_{\lambda,\gamma}(A_1^p, C_1^r)$$

**7th Case**: if $\begin{cases} \delta_{\lambda,\gamma}(A_1^p, B_1^q) = \delta_{\lambda,\gamma}(A_1^{p-1}, B_1^{q-1}) + \Gamma(a'_p \to b'_q) \\ \delta_{\lambda,\gamma}(B_1^q, C_1^r) = \delta_{\lambda,\gamma}(B_1^{q-1}, C_1^r) + \Gamma(b'_q \to \Lambda) \end{cases}$, then

$$\Sigma = \delta_{\lambda,\gamma}(A_1^{p-1}, B_1^{q-1}) + d(a'_p, b'_q) + d(a'_{p-1}, b'_{q-1}) + \delta_{\lambda,\gamma}(B_1^{q-1}, C_1^r) + d(b'_q, b'_{q-1}) + \lambda$$
$$\geq \delta_{\lambda,\gamma}(A_1^{p-1}, B_1^{q-1}) + \delta_{\lambda,\gamma}(B_1^{q-1}, C_1^r) + d(a'_p, a'_{p-1}) + \lambda \qquad (d \text{ satisfies the triangle inequality})$$
$$\geq \delta_{\lambda,\gamma}(A_1^{p-1}, C_1^r) + d(a'_p, a'_{p-1}) + \lambda \underset{def}{\geq} \delta_{\lambda,\gamma}(A_1^p, C_1^r) \qquad (H4) \text{ applies}.$$

**8th Case**: if $\begin{cases} \delta_{\lambda,\gamma}(A_1^p, B_1^q) = \delta_{\lambda,\gamma}(A_1^{p-1}, B_1^q) + \Gamma(a'_p \to \Lambda) \\ \delta_{\lambda,\gamma}(B_1^q, C_1^r) = \delta_{\lambda,\gamma}(B_1^{q-1}, C_1^r) + \Gamma(b'_q \to \Lambda) \end{cases}$, then:

$$\Sigma = \delta_{\lambda,\gamma}(A_1^{p-1}, B_1^q) + d(a'_p, a'_{p-1}) + \lambda + \delta_{\lambda,\gamma}(B_1^q, C_1^r)$$
$$\geq \delta_{\lambda,\gamma}(A_1^{p-1}, C_1^r) + d(a'_p, a'_{p-1}) + \lambda, \quad (H4) \text{ applies}$$
$$\underset{def}{\geq} \delta_{\lambda,\gamma}(A_1^p, C_1^r)$$

**9th Case**: if $\begin{cases} \delta_{\lambda,\gamma}(A_1^p, B_1^q) = \delta_{\lambda,\gamma}(A_1^p, B_1^{q-1}) + \Gamma(\Lambda \to b'_q) \\ \delta_{\lambda,\gamma}(B_1^q, C_1^r) = \delta_{\lambda,\gamma}(B_1^{q-1}, C_1^r) + \Gamma(b'_q \to \Lambda) \end{cases}$, then:

$$\Sigma \geq \delta_{\lambda,\gamma}(A_1^p, B_1^{q-1}) + \delta_{\lambda,\gamma}(B_1^{q-1}, C_1^r)$$
$$\geq \delta_{\lambda,\gamma}(A_1^p, C_1^r) \quad (H4) \text{ applies}.$$

So property *P4* holds for all *m* in $\{0,..n\}$. By induction *P4* holds for all m in *N* and so *P4* holds for all $(A_1^p, B_1^q, C_1^r)$ in $U \times U \times U$. □

# IV. Proposition 2: $\delta_{\lambda,\gamma}$ is upper bounded by twice the *Lp* distance.

*Proposition 2:*

$\forall \lambda \geq 0, \gamma > 0 \quad \forall X, Y \in U^2 \quad \delta_{\lambda,\gamma}(X,Y) \leq 2 \cdot D_{LP}(X,Y)$, whenever *X* and *Y* have the same length.

Proof: let us consider the sequence of editing operations consisting in *m* match operations, where *m* is the length of the *X* and *Y*. This sequence has a cost equal to twice the *Lp-distance* between the two time series *X* and *Y*. Since $\delta_{\lambda,\gamma}$ is equal to the cost of the optimal sequence of edit operations, the result follows. □

# V. Proposition 3: $\delta_{\lambda,\gamma}$ is an increasing function of $\lambda$ and $\gamma$

*Proposition 3:*

$$\forall \lambda \geq 0, \gamma > 0 \; \forall \lambda' \geq \lambda \; \forall \gamma' \geq \gamma \; \forall X,Y \in U^2 \quad \delta_{\lambda,\gamma}(X,Y) \leq \delta_{\lambda',\gamma'}(X,Y)$$

Proof: Let us consider one of the optimal sequences of editing operations evaluated with the tuple $(\lambda', \gamma')$ with minimal cost equal to $\delta_{\lambda',\gamma'}(X,Y)$. If we keep this sequence of editing operation while replacing $(\lambda', \gamma')$ with $(\lambda, \gamma)$ in all the elementary operation costs we get a cost for this sequence that is lower than $\delta_{\lambda',\gamma'}(X,Y)$ but greater than the cost of the optimal sequence $\delta_{\lambda,\gamma}(X,Y)$ evaluated using $(\lambda, \gamma)$. The result follows. □

# VI. Proposition 4: Upper-bounding the distance between a time series and its piecewise constant polygonal approximation.

We define $\overline{A}_1^{p,r}$ as a Piece Wise Constant Approximation (PWCA) of time series $A_1^p$ containing $r-1 \geq 0$ constant segments and $p$ samples. This approximation can be obtained using any kind of solution (from heuristic to optimal solutions), let say the optimal solution similar to the one proposed in [8]. $\overline{A}_1^{p,r}$ and $A_1^p$ have the same number of samples, namely $p$. Let $\tilde{A}_1^r$ be the time series composed with the $r$ segment extremities of $\overline{A}_1^{p,r}$. $\tilde{A}_1^r$ contains $r$ samples. Let us similarly define $\overline{B}_1^{p,r'}$ and $\tilde{B}_1^{r'}$ from time series $B_1^p$.

*Proposition 4:*

$$\forall \lambda \geq 0, \gamma > 0, \quad \forall r \in [1; p[, \quad \forall X_1^p \in U \quad \delta_{\lambda,\gamma}(\overline{X}_1^{p,r}, \tilde{X}_1^r) \leq \lambda \cdot (p-r) + \gamma \cdot \Delta T(2 \cdot p - r) \quad ,$$

where $\Delta T$ is the time difference average between two successive samples inside the piecewise constant segments of the approximation.

The proof for this proposition is straightforward: let us consider the sequence of operations consisting in $r$ match operations for the end extremities of the piecewise constant segments and *(p-r)* delete operations for the set of samples in $\overline{X}_1^{p,r}$ that are not end extremities of the piecewise constant segments. In this sequence each match operation has in average the cost $(p/r) \cdot \gamma \cdot \Delta T$, and each delete operation has a $\lambda$ fixed penalty and a penalty proportional to the time stamp difference between two successive samples $\gamma \cdot (timeStamps(i) - timeStamps(i-1))$. Then, the cost for this sequence of editing operations is $(p-r)(\lambda + \gamma \cdot \Delta T) + p \cdot \gamma \cdot \Delta T$. Finally the optimal sequence of editing operations has a cost $\delta_{\lambda,\gamma}(\overline{X}_1^{p,r}, \tilde{X}_1^r)$ lower or equal to $\lambda \cdot (p-r) + \gamma \cdot \Delta T(2 \cdot p - r)$. □

## VII Conclusion

We have proposed a family of time warp edit distances for time series matching. This family involves two parameters: the *stiffness* parameter that controls the elasticity of the distance and the *gap penalty* that is part of the cost involved in the *insert* or *delete* operations. We have shown that the proposed measure:

- is a metric on the set of discrete and finite time series.
- is upper bounded by twice the *Lp-distance*.
- is an increasing function of the *gap penalty* and the *stiffness* parameter.

Further more, the distance between a time series and its piecewise constant approximation is upper bounded by an expression that only depends on the lengths of the times series, the number of segments of its approximation and the two parameters $\lambda$ and $\gamma$.